\documentclass[aps,pra,superscriptaddress,twocolumn,longbibliography]{revtex4-1}
\usepackage[utf8]{inputenc}
\usepackage{hyperref}
\usepackage{amsmath}
\usepackage{amsfonts}
\usepackage{comment}
\usepackage{quantikz}

\newsavebox{\boxA}
\newsavebox{\boxB}
\newsavebox{\boxC}

\usepackage[caption=false]{subfig}

\usepackage{geometry}
\usepackage{color}

\hypersetup{hidelinks,colorlinks=true,breaklinks=true,urlcolor=blue,allcolors=blue,pdftitle={Title},pdfauthor={Author}}

\begin{document}

\title{Experimental high-dimensional Greenberger-Horne-Zeilinger entanglement\\with superconducting transmon qutrits}
\author{Alba Cervera-Lierta}
\email{a.cervera.lierta@gmail.com}
\affiliation{Chemical Physics Theory Group, Department of Chemistry, University of Toronto, Canada}
\affiliation{Department of Computer Science, University of Toronto, Canada}
\affiliation{Barcelona Supercomputing Center}
\author{Mario Krenn} 
\email{mario.krenn@mpl.mpg.de}
\affiliation{Chemical Physics Theory Group, Department of Chemistry, University of Toronto, Canada}
\affiliation{Department of Computer Science, University of Toronto, Canada}
\affiliation{Vector Institute for Artificial Intelligence, Toronto, Canada}
\affiliation{Max Planck Institute for the Science of Light (MPL), Erlangen, Germany}
\author{Al\'an Aspuru-Guzik}
\email{alan@aspuru.com}
\affiliation{Chemical Physics Theory Group, Department of Chemistry, University of Toronto, Canada}
\affiliation{Department of Computer Science, University of Toronto, Canada}
\affiliation{Vector Institute for Artificial Intelligence, Toronto, Canada}
\affiliation{Canadian  Institute  for  Advanced  Research (CIFAR)  Lebovic  Fellow,  Toronto,  Canada}
\author{Alexey Galda}
\email{agalda@uchicago.edu}
\affiliation{Menten AI, Inc., San Francisco, CA 94111, USA}
\affiliation{James Franck Institute, University of Chicago, Chicago, IL 60637, USA}
\affiliation{Computational Science Division, Argonne National Laboratory, Lemont, IL 60439, USA}

\begin{abstract}
Multipartite entanglement is one of the core concepts in quantum information science with broad applications that span from condensed matter physics to quantum physics foundations tests. Although its most studied and tested forms encompass two-dimensional systems, current quantum platforms technically allow the manipulation of additional quantum levels.
We report the experimental demonstration and certification of a high-dimensional multipartite entangled state in a superconducting quantum processor. We generate the three-qutrit Greenberger-Horne-Zeilinger state by designing the necessary pulses to perform high-dimensional quantum operations. 
We obtain the fidelity of $76\pm 1\%$, proving the generation of a genuine three-partite and three-dimensional entangled state.
To this date, only photonic devices have been able to create and certify the entanglement of these high-dimensional states.
Our work demonstrates that another platform, superconducting systems, is ready to exploit genuine high-dimensional entanglement and that a programmable quantum device accessed on the cloud can be used to design and execute experiments beyond binary quantum computation.
\end{abstract}

\maketitle

\section{Introduction}

Entanglement is one of the most striking properties of quantum mechanics that has fascinated the physics community for the last 80 years. In the words of Schr\"odinger, he ``would not call that \textit{one} but rather \textit{the} characteristic trait of quantum mechanics, the one that enforces its entire departure from classical lines of thought" \cite{schrodinger1935discussion}. The pioneer studies of quantum entanglement were restricted to two-particle phenomena, the well-known Einstein–Podolsky–Rosen (EPR) pairs \cite{einstein1935can}. 
Multipartite entangled states present richer physical implications and a more complex mathematical structure.
The main representative of a \textit{genuine} multipartite entangled state was proposed by Greenberger-Horne-Zeilinger (GHZ) to show that local realism can be violated by quantum mechanics with deterministic experiments \cite{greenberger1989going,pan2000experimental}. Since then, multipartite entanglement has become a key topic in quantum information science for its rich and challenging mathematical structure \cite{bengtsson2017geometry}, its implications in the study of local-realism theories \cite{brunner2014bell}, its use as a resource in quantum sensing protocols \cite{giovannetti2011advances,quantumsensing,liu2021distributed} or the study of critical phases in condensed matter models \cite{osterloh2002scaling,vidal2003entanglement}.

The vast majority of experimental entanglement studies have focused on two-dimensional systems. However, high-dimensional entanglement is broadly present in nature and can experimentally be generated in our labs. High-dimensional states have been extensively used in photonic systems \cite{kues2017chip, erhard2020advances, wang2020integrated}. Other experimental platforms such as trapped ions have simulated the dynamics of spin-1 systems \cite{senko2015realization}, while ensembles of cold atoms can be used as quantum memories to store high-dimensional states \cite{parigi2015storage,ding2016high}. 
Nevertheless, to our knowledge, only photonic systems have been able to generate and perform quantum information tasks that require high-dimensional multipartite entanglement \cite{erhard2020advances},  albeit the work from Ref.\cite{blok2020quantum} requires theoretically high-dimensional entanglement, although it is not experimentally certified that is genuinely multi-partite.

Quantum control beyond the two-level system has been exploited in superconducting quantum processors since the very beginning of this technology. Starting from the use of the higher levels for qubit readout \cite{martinis2002rabi,cooper2005observation,lucero2008high}, fast qubit initialization \cite{valenzuela2006microwave, reed2010fast,geerlings2013demonstrating}, and the explicit use of the third level for spin-1 quantum simulation \cite{neeley2009emulation}, the first steps towards the realization of ternary quantum computation with superconducting transmon devices have been taken in the last 10 years \cite{bianchetti2010control,abdumalikov2010electromagnetically,abdumalikov2013experimental,bianchetti2010control,vepsalainen2020simulating,tan2018topological, honigl2018mixing,jerger2016contextuality,neeley2009emulation, fedorov2012implementation}. More recently, these efforts have led to the implementation of high-fidelity single-qutrit gates \cite{yurtalan2020implementation, morvan2020qutrit}. In addition to paving the way to extend quantum computation to the three- and higher-level systems, these experiments enable the study of high-dimensional quantum physics.

\begin{figure}[t!]
    \centering
    \includegraphics[width=\columnwidth]{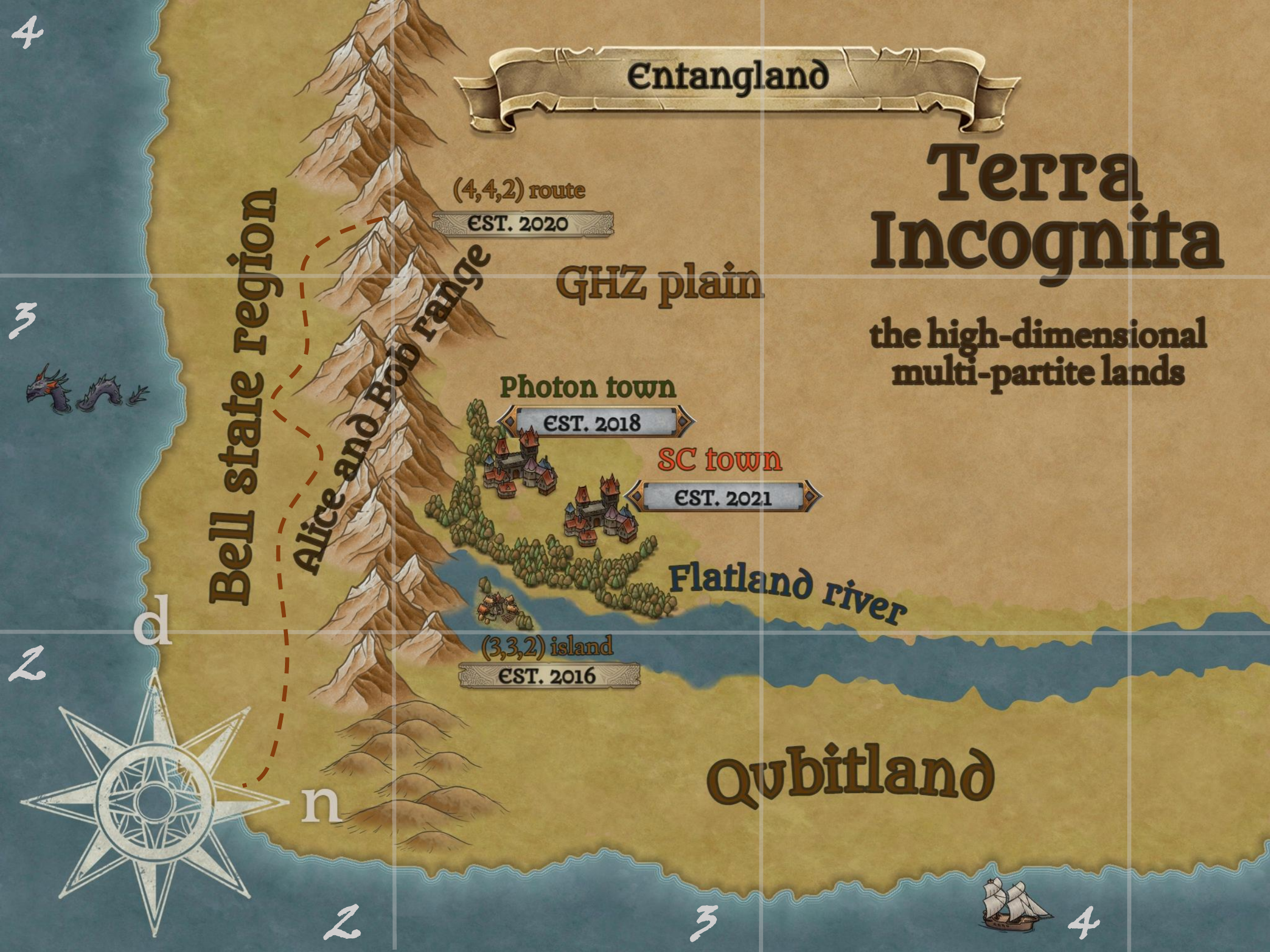}
    \caption{Map of \emph{Entangland}. It represents the experimental state-of-the-art generation of high-dimensional multipartite states. Bipartite entangled states have been generated for up to 100 dimensions with photons \cite{romero2012increasing,krenn2014generation}, and recently a qutrit Bell state was generated with superconducting circuits \cite{blok2020quantum} (\textit{Bell state region}). On the other hand, multipartite qubit states have also been produced in different platforms \cite{PhysRevLett.119.180511,zhong201812,wang201818, song2019generation,omran2019generation, pogorelov2021compact, gao2019entanglement} (\textit{Qubitland}). The present work demonstrates the experimental generation and certification of a three-dimensional GHZ state with superconducting quantum devices. There are only two photonic experiments that have been able to explore Entangland beyond \textit{Alice and Bob range}, i.e. the divide between bipartite and multipartite entanglement, and to cross the \textit{Flatland river} \cite{malik2016multi, erhard2018experimental,hu2020experimental}, i.e. the divide between qubit states and $d>2$ level subsystems. Besides these photonic works, the experiment from Ref.\cite{blok2020quantum} requires high-dimensional entanglement, although it is not certified the particular amount experimentally generated.}
    \label{fig:map}
\end{figure}

High dimensional multipartite entanglement has never been observed and certified outside of photonic systems before. Here we show and certify the generation of a high-dimensional GHZ state in a superconducting quantum system, i.e. the equal superposition state $(|000\rangle + |111\rangle + |222\rangle)/\sqrt{3}$. Our experiment is run in a programmable device at least 30000 times faster than the preceding photonic experiment \cite{erhard2018experimental}. The ability to generate and certify the qutrit GHZ state in a programmable cloud-based quantum processor offers a pathway to investigate the almost unexplored world of multiparticle high-dimensional entanglement, as shown pictorially in Fig. \ref{fig:map}. The experiment is executed in the five-transmon IBM Quantum system \texttt{ibmq\_rome} using Qiskit Pulse \cite{alexander2020qiskit} to program and calibrate the single-qutrit gate set in the $(12)$ subspace. We report $76 \pm 1\%$ fidelity, exceeding the entanglement witness threshold to demonstrate genuine high-dimensional multipartite entanglement. Our results, together with the recent results from Ref.\cite{blok2020quantum}, show how to explore multi-level entanglement with a non-photonic experiment.

\section{Experiment}

\begin{figure*}[t!]
    \centering
\includegraphics[width=1\textwidth]{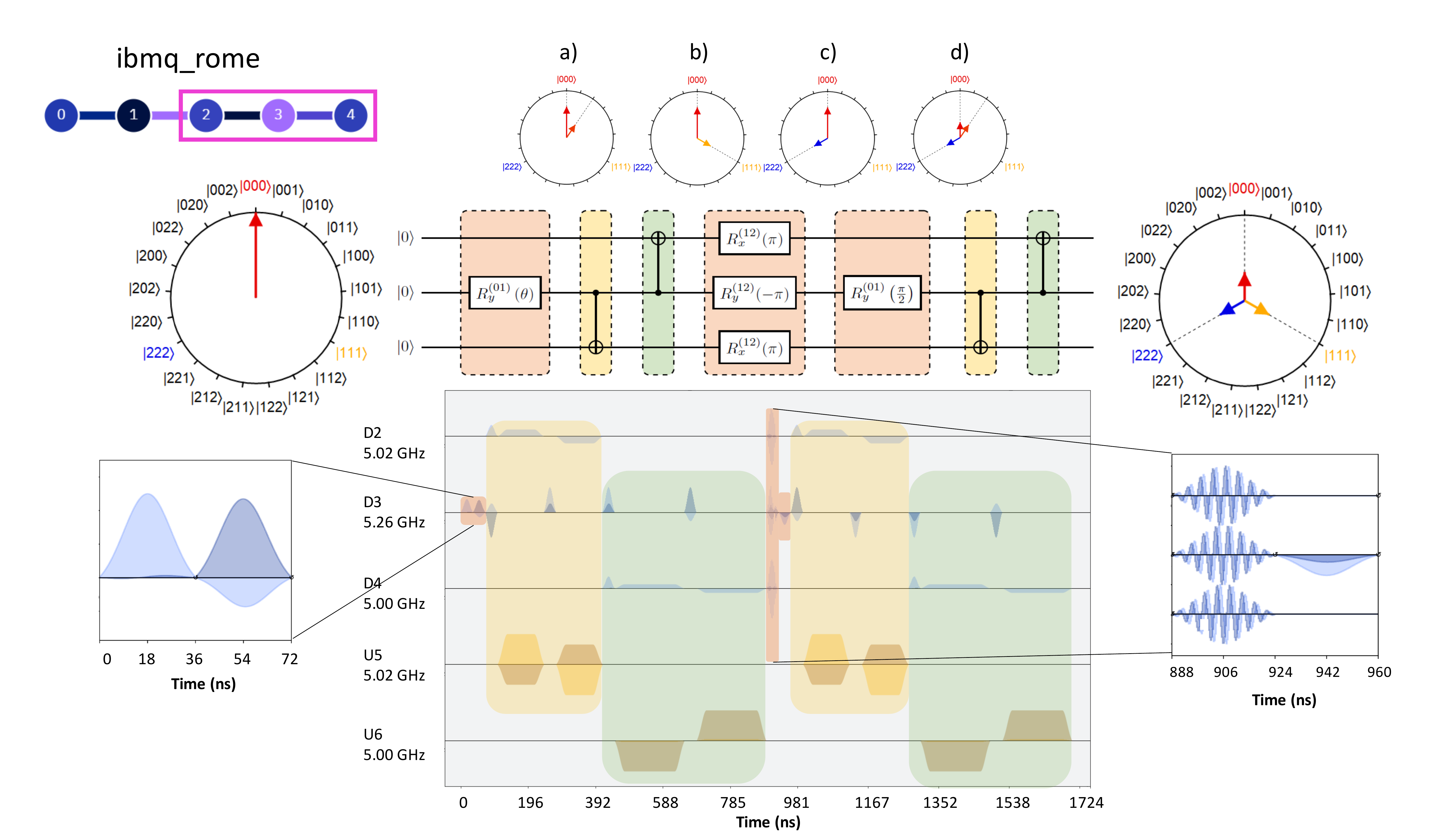}
    \caption{Quantum circuit to generate the qutrit GHZ state, its diagrammatic representation through the $GHZ-clock$, and the pulse schedule. The single-qutrit gates act in the $(01)$ and $(12)$ subspaces, and the CNOT gates act as described in Tab. \ref{tab:gatesb}. (a) First, a superposition in the $(01)$ subspace is generated in the second qutrit using a $R^{(01)}_{y}$ gate with angle $\theta=2\,\text{arctan}\left(1/\sqrt{2}\right)$. (b) Next, two CNOT gates are applied to create two-dimensional entanglement across all three transmons. (c) The superposition state is shifted from the $(01)$ to the $(12)$ subspace by using $R_{x}^{(12)}$ and $R_{y}^{(12)}$ gates. (d) The third basis element is created by generating another $(01)$ superposition, followed by repeating the application of CNOT gates controlled on the central qutrit to finish the qutrit GHZ state. At the bottom, the pulse schedule that executes the circuit. Drive channels d1, d2, and d3 are used to implement calibrated microwave pulses of qutrits Q2, Q3, and Q4 from the \texttt{ibmq\_rome} device, respectively. The cross resonance gates utilized to implement the CNOT operations are executed on channels U2 and U5 by transmitting two-transmon control drives.
    }
    \label{fig:GHZ}
\end{figure*}

In the next subsections, we describe the different steps followed to generate the three-dimensional GHZ state with superconducting transmon qutrits.

\subsection{Qutrit gates and measurement protocol}

The experiment was performed using the IBM Quantum cloud provider employing the set of default single- and two-qubit gates, augmented by our proposed calibrated single-qutrit gate set. We used the Qiskit Pulse \cite{alexander2020qiskit}, a pulse-level programming model that allows us to define, calibrate and execute quantum circuits. The users can either use qubit gates calibrated by the IBM team or design their own microwave pulses. As a consequence, they can perform a wide set of quantum operations on the transmon chip. 

Such low-level access to quantum hardware enables encoding and processing quantum information in qutrits (three-level systems), extending the concept of quantum computation beyond two-level systems.

The local oscillator (LO) frequency for each transmon in the device, given by the calibrated $|0\rangle \rightarrow |1\rangle$ frequency, was kept fixed in the experiment. Transitions between levels $|1\rangle$ and $|2\rangle$ were implemented using amplitude-modulated microwave pulses by applying a sinusoidal side-band at the frequency $f_{12} - f_{01}$, 
effectively shifting the frequency of the pulses from $f_{01}$ to $f_{12}$.

To design the quantum circuit that generates the qutrit GHZ state we will use a combination of default single-qubit gates, labeled with a superscript $(01)$, entangling CNOT gate, and custom calibrated single-qutrit gates acting in the $(12)$ subspace (see \ref{tab:gatesa}) $R_{\alpha}^{(12)}\left(\theta\right)=e^{-i\frac{\theta}{2}\sigma_{\alpha}^{(12)}}$,
where $\alpha=x,y$, and $\sigma_{\alpha}^{(12)}$ are the Pauli matrices $\sigma_{\alpha}$ acting in the $(12)$ subspace. We perform a Rabi experiment to obtain the amplitude of the $\pi_{1\rightarrow 2}$ pulse and use it to define these rotational gates.
Default hardware implementation of the CNOT gate was used, as defined by IBM Quantum. Extended to qutrits, it acts as a $SU(9)$ gate with the truth table shown in Tab. \ref{tab:gatesb}. For the control qutrit in the $(01)$ subspace, it acts as a standard qubit CNOT and adds a $\pi/2$ phase to the $|2\rangle$ state of the target qutrit. When the control qutrit is in the $|2\rangle$ state, it generates an equal superposition in the $(01)$ space of the target qutrit and adds a relative phase to the $|2\rangle$ state. The full characterization of the action of a CNOT gate in the $SU(9)$ space is provided in Ref.\cite{galda2021implementing}.

\savebox{\boxA}{\begin{quantikz} & \gate{R_{y}^{(01)}\left(\theta\right)} & \qw \end{quantikz}}%
\savebox{\boxB}{\begin{quantikz} & \gate{R_{y}^{(12)}\left(\theta\right)} & \qw \end{quantikz}}%
\savebox{\boxC}{\begin{quantikz} & \gate{R_{x}^{(12)}\left(\theta\right)} & \qw \end{quantikz}}%
\begin{table*}[]

    \subfloat[Single-qutrit gates.]{\begin{tabular}{ccc}
    Gate & Matrix\\
    \hline
    \hline
    \usebox\boxA &  $\left(\begin{matrix}
              \cos(\theta/2) & -\sin(\theta/2) & 0  \\
              \sin(\theta/2) & \cos(\theta/2) & 0 \\
              0 & 0 & 1
         \end{matrix}\right)$\\
     \usebox\boxB &  $\left(\begin{matrix}
              1 & 0 & 0 \\
              0 & \cos(\theta/2) & -\sin(\theta/2)  \\
              0 & \sin(\theta/2) & \cos(\theta/2)              
         \end{matrix}\right)$\\   
     \usebox\boxC &  $\left(\begin{matrix}
              1 & 0 & 0 \\
              0 & \cos(\theta/2) & -i\sin(\theta/2)  \\
              0 & -i\sin(\theta/2) & \cos(\theta/2)            
         \end{matrix}\right)$
    \end{tabular}
        \label{tab:gatesa}
    }%
    \hspace{1.3cm}
    \subfloat[Truth table IBM default CNOT gate.]{\begin{tabular}{ccc}
       Control & Target & Output \\
       \hline
       \hline
       $|0\rangle$ & $|0\rangle$ & $|00\rangle$ \\
       $|0\rangle$ & $|1\rangle$ & $|01\rangle$ \\
       $|0\rangle$ & $|2\rangle$ & $|02\rangle$ \\
       $|1\rangle$ & $|0\rangle$ & $|10\rangle$ \\
       $|1\rangle$ & $|1\rangle$ & $|11\rangle$ \\
       $|1\rangle$ & $|2\rangle$ & $i|12\rangle$ \\
       $|2\rangle$ & $|0\rangle$ & $a|20\rangle+b|21\rangle$ \\
       $|2\rangle$ & $|1\rangle$ & $b^{*}|20\rangle+c|21\rangle$ \\
       $|2\rangle$ & $|2\rangle$ & $e^{i\varphi}|22\rangle$ 
    \end{tabular}
        \label{tab:gatesb}
    }
        \caption{Ternary quantum gates used to generate the GHZ state. We use the default pulse calibration for the qubit gates $R_{y}^{(01)}$ and CNOT and design the pulse sequence for the qutrit gates $R_{y}^{(12)}$ and $R_{x}^{(12)}$. For the CNOT gate, one needs to take into account its effect when the control is in the $|2\rangle$ state. This effect is reproduced in the CNOT truth table shown (more details can be found in Ref. \cite{galda2021implementing}).}
    \label{tab:gates}
\end{table*}

IBM Quantum provides the device specifications after each calibration cycle, as well as a basic set of quantum gates that include single and two-qubit gates. The device specifications for \texttt{ibmq\_rome} used in the experiment are shown in Tab.~\ref{Table:chip_properties}.
The qutrit GHZ state was experimentally achieved on transmons Q2, Q3, and Q4 connected in a linear chain.

As a qutrits readout protocol, we adopted the default $0-1$ state discriminator, as implemented by IBM Quantum, to most accurately distinguish between the $|000\rangle$ state and the rest of possible computational basis states. 
This discriminator is unable to correctly identify excitations to the $|2\rangle$ state, misclassifying them as $|1\rangle$. As a result, in order to measure other basis states, we perform local operations to lower the qutrits to the $|000\rangle$ state. For instance, to measure the probability of obtaining the $|010\rangle$ state, we apply $X^{(01)}$ on the second qutrit and measure the probability of the $|000\rangle$ state. Although this protocol increases the number of necessary experiments (one for each computational basis probability amplitude), it is aimed at reducing measurement errors by taking advantage of the high accuracy of the proprietary $0-1$ discriminator offered by IBM Quantum, eliminating the need to calibrate a custom 0-1-2 discriminator, and only measuring the state with the lowest readout error. 

In Fig.~\ref{fig:mit} from App. \ref{app:chip}, we present the measurement error mitigation matrix, commonly used to mitigate state preparation and measurement (SPAM) errors. It is obtained by preparing each computational basis state and collecting the probabilities of measuring each of these states. The errors obtained in that matrix resembles the readout errors provided by the IBM team, which indicates that we are essentially correcting readout errors. This conclusion is extracted under the assumption that the readout errors provided by IBM are indeed only readout errors, and that initialization errors are characterized and negligible in comparison.

The bottom panel of Fig.~\ref{fig:GHZ} shows the pulse sequence for the quantum circuit executed to create the three-dimensional GHZ state. Local qutrit gates were implemented by transmitting calibrated microwave pulses to manipulate qutrits Q2, Q3 and Q4 through the drive channels D2, D3, and D4, correspondingly. Cross-resonance-based CNOT gates between transmons Q2 and Q3, and between Q2 and Q4 were implemented by transmitting two-transmon control drives on channels U5 and U6, correspondingly, with rotary target corrections~\cite{sundaresan2020reducing}. 

\subsection{The GHZ circuit}

The three-dimensional GHZ state in the computational basis can be written as
\begin{equation}
    |GHZ\rangle = \frac{1}{\sqrt{3}}\left(|000\rangle + |111\rangle + |222\rangle\right).
    \label{eq:GHZ}
\end{equation}
This state is a maximally entangled state with Schmidt rank vector of $(3,3,3)$ \cite{huber2013structure} and maximal entropy for all three bipartitions. It is also a genuine three-partite and three-dimensional entangled state, i.e. the entanglement is distributed into the three parties in a way it cannot be contained in other entangled states like the Bell states (see App.\ref{app:entanglement} for details).

With the experimental gate set presented in the previous subsection, we adapted the \textsc{Melvin} algorithm \cite{krenn2016automated,krenn2020computer} (an automated design algorithm for quantum optics that learns over time to find experiments faster by improving its own toolbox) to operate with digital quantum gates. We restricted the angles of the rotation gates to $\theta=\pm \pi$ and $\theta=\pm \pi/2$ (except for only up to one free angle for the $R_{y}^{(01)}$ gate) to look for a solution that, besides being practical, can also be interpreted. Indeed, \textsc{Melvin}'s solution dealt with the inconvenient relative phases introduced by the CNOT gate, respected the chip connectivity (and thus, used the minimal number of available gates) while provided us with an interpretable circuit. 

The circuit found to generate the GHZ state is shown in Fig. \ref{fig:GHZ}. It requires four CNOT gates, two single-qubit gates and three single-qutrit gates.
The first part of the circuit only acts in the $(01)$ subspace while in the second part both control and target qutrits are in the $|2\rangle$ state and, thus, they do not generate any superposition. We can only expect a relative phase $2\varphi$ to appear in the $|222\rangle$ state, which does not change the entanglement properties of the GHZ state generated. To illustrate how the quantum circuit produces the GHZ state, we introduce the \emph{GHZ-clock} representation, briefly described in the App. \ref{app:GHZ_clock}.

\subsection{Fidelity and entanglement witness protocol}

To obtain the fidelity of the generated state with respect to the theoretical GHZ state from Eq. \eqref{eq:GHZ}, we follow the steps presented in Refs.~\cite{bavaresco2018measurements,erhard2018experimental}. Notice that we do not require to perform a full tomography protocol to obtain the fidelity. Denoting $\rho$ as the density matrix from the state generated in the chip, the fidelity with respect to the GHZ state can be computed as $\text{Tr}\left(\rho|GHZ\rangle\langle GHZ|\right)$, i.e.
\begin{equation}
    F_\mathrm{exp} = \frac{1}{3}\bigg(\sum_{i=0}^{2}\langle iii|\rho|iii\rangle + 2 \sum_{\substack{i,j=0\\i<j}}^{2} \text{Re}\langle iii|\rho|jjj\rangle\bigg).
    \label{eq:fidelity}
\end{equation}
The first sum corresponds to measure the probability amplitudes of the computational basis states $|000\rangle$, $|111\rangle$ and $|222\rangle$. Due to the Hermiticity of the density matrix, the second sum contains only three independent terms, $\langle 000|\rho|111\rangle$, $\langle 000|\rho|222\rangle$ and $\langle 111|\rho|222\rangle$. The real and imaginary parts of these terms can be measured by computing the expectation value of certain combinations of $\sigma_{x}$ and $\sigma_{y}$ operators \cite{erhard2018experimental} (see App. \ref{app:tomography} for details), although we will only need the real part to obtain the fidelity.

The common approach to implementing quantum gates on IBMQ quantum processors consists in using a single rotating reference frame associated with the $f_{01}$ qubit frequency \cite{krantz2019quantum}. When implementing single-qutrit gates in the (12) subspace by using amplitude-modulated pulses, an additional phase is accumulated in the (01) subspace, and is proportional to the gate duration and the transmon anharmonicity. This phase can be compensated by applying an additional $R_{z}^{(01)}$ gate or by introducing a delay before the implementation of these gates.
During the generation of the GHZ state, this phase appears as a relative phase in the $|222\rangle$ basis state and, therefore, it does not affect the entanglement properties of the state. However, during the tomography protocol described above, these phases will become relevant and affect the result if they are not properly canceled. To do so, we apply a $R_{z}^{(01)}(\varphi)$ gate in between the GHZ generation circuit and the tomography gates and scan for different values of $\varphi$. Then, we observe the expected oscillations in the probability amplitudes and choose the results according to the $\varphi$ that has compensated the phase accumulation.
More details about this analysis can be found in App. \ref{app:track_phase}.

The final step is to certify the generation of three-partite entanglement. To do so, we follow the entanglement witness protocol described in Ref. \cite{malik2016multi}. This method establishes that the maximum possible fidelity between a GHZ state and an immediately lower-dimensional entangled state (i.e. a state with Schmidt rank vector of $(3,3,2)$, $(2,3,3)$ or $(3,2,3)$) is $2/3$. Thus, the condition $F_\mathrm{exp}>2/3$ certifies the generation of a three-dimensional three-partite entangled state. More details about this protocol are presented in App. \ref{app:tomography}.

\section{Results and discussion}

\begin{figure*}[t!]
    \centering
    \includegraphics[width=1\textwidth]{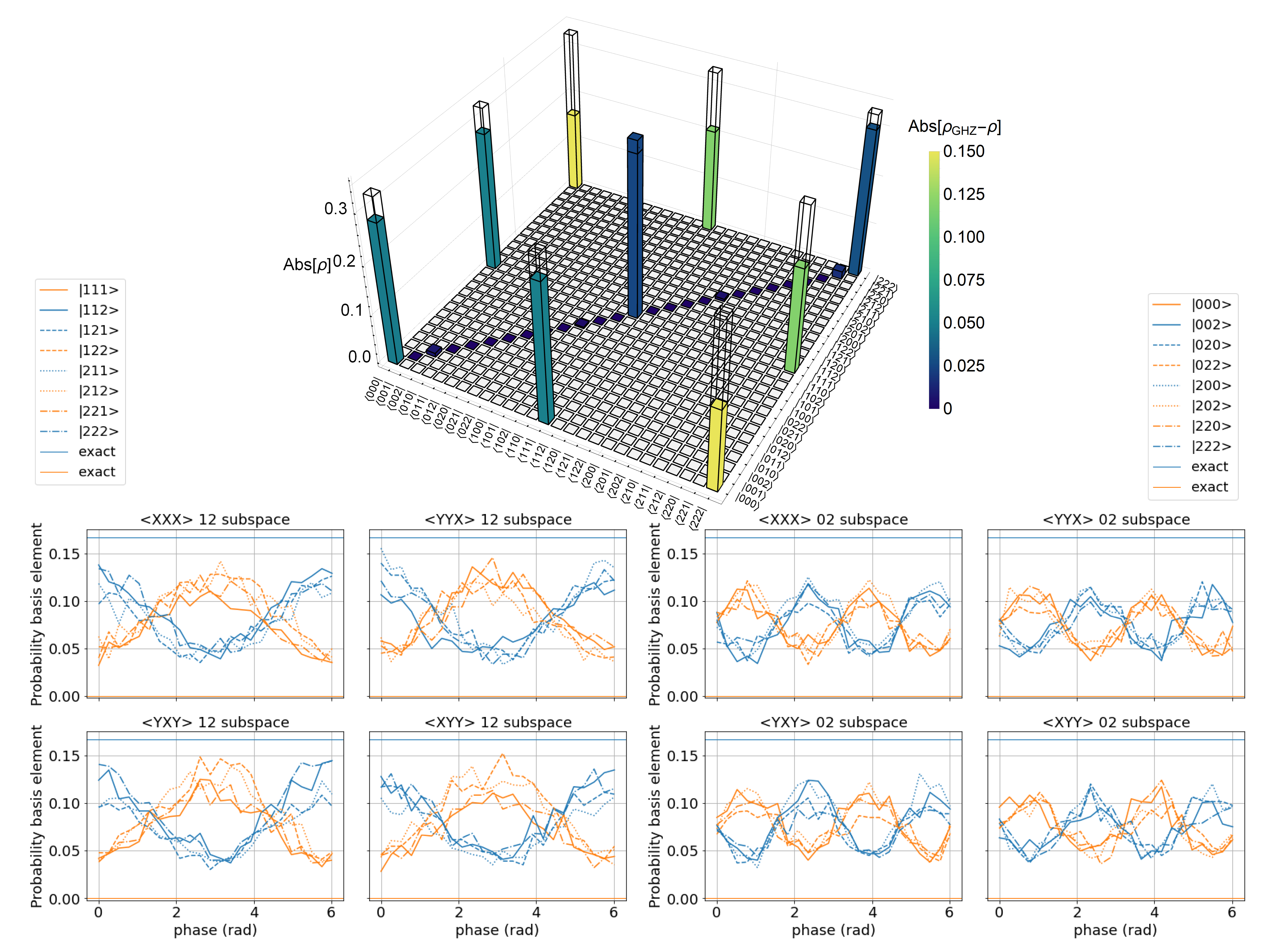}\\
    \caption{Top: experimental results of the density matrix elements needed to compute the fidelity with respect to the GHZ state. Only six elements are required to reconstruct the fidelity: the three diagonal terms $\langle 000|\rho|000\rangle$, $\langle 111|\rho|111\rangle$ and $\langle 222|\rho|222\rangle$, and the off-diagonal terms $\langle 000|\rho|111\rangle$, $\langle 000|\rho|222\rangle$ and $\langle 111|\rho|222\rangle$. The plot shows the absolute value of these elements (in grey, all other elements not measured). The corresponding fidelity obtained is $76 \pm 1\%$, exceeding the 2/3 threshold required to certify genuine three-partite entanglement. The real parts of these elements are presented in App.~\ref{app:error_data}. Bottom: probability amplitudes of each computational basis element as a function of the phase introduced with a $R_{z}^{(01)}$ between the GHZ generation and the gates used for the tomography. As expected, these amplitudes oscillate due to the reference frame imposed in the (01) subspace.}
    \label{fig:fidelity}
\end{figure*}

We performed a total of 1595 experiments to reconstruct the density matrix elements necessary to compute the fidelity with respect to the GHZ state (Eq.~\eqref{eq:fidelity}).
For the (01) subspace, we need to measure four expectation values corresponding to the $X$ and $Y$ projective measurements. Each of these expectation values requires 8 different circuits to obtain the result by measuring only the $|000\rangle$ state, i.e. a total of $4\cdot 8=32$ experiments. We repeat this process for the (12) and (02) subspace but, in those cases, we need to find the proper phase that cancels the precession effect explained in the previous section. To do so, we scan for different phases between $[0, 2\pi]$ in intervals of $\pi/12$ for each expectation value. This makes a total of $2\cdot 4\cdot 8 \cdot 24 = 1536$ experiments for these two subspaces. Finally, we also measure the 27 diagonal elements.
The number of shots for each experiment was $n=1024$.

The reconstruction of the density matrix is depicted in Fig. \ref{fig:fidelity}. It does not include all off-diagonal terms needed to reconstruct entirely the quantum state, but only those necessary to calculate the fidelity with respect to the GHZ state. The real part of each of these elements is shown in Eq. \eqref{eq:density_matrix_real} from App. \ref{app:error_data}.
We obtain a fidelity of
\begin{equation}
    F_\mathrm{exp}= 0.76 \pm 0.01,
    \label{eq:fideltiy_result}
\end{equation}
after applying the readout error mitigation, and $F_\mathrm{raw}=0.69 \pm 0.01$ with the raw measured data, in both cases, exceeding the bound of 2/3 required to certify three-dimensional three-partite entanglement.
The trace of the density matrix is 1.12 for the raw data and 0.98 for the mitigated data. This shows that the error mitigated results are more reliable than the raw ones, implying that a significant source of error comes from the incorrect basis state assignment after performing the measurement.

In this particular experiment, we only reconstruct the real part of the density matrix elements. The error with respect to the ideal 1/3 in the off-diagonal elements, could suggest possible relative phases between the three GHZ computational basis states. Notice that the state $(|000\rangle + e^{i\phi_{1}}|111\rangle + e^{i\phi_{2}}|222\rangle)/\sqrt{3}$ with $-\pi\leq\phi_{1},\phi_{2}\leq\pi$, has the same entanglement properties as the GHZ from Eq. \eqref{eq:GHZ}. We performed another experiment to measure the imaginary part of these elements that allow us to obtain these relative phases. The fidelity obtained in that experiment is $F_\mathrm{exp}=0.78\pm 0.01$ and $F_\mathrm{raw}=0.69\pm 0.02$ and the relative phases obtained are $\phi_{1}=-0.04 \pm 0.03$ and  $\phi_{2} = 0.25 \pm 0.03$. More details about this second experiment are provided in App. \ref{app:relative_phases}.

Various sources of errors can affect the fidelity of the GHZ state. Working with higher-dimensional states opens up additional channels of decoherence, including $|2\rangle\rightarrow|1\rangle$ relaxation (T1 process) and dephasing (T2 process) in the (12) subspace, in addition to these decoherence channels in the (01) qubit subspace. We also expect that the errors coming from qubit crosstalk, which are common in superconducting transmon devices, can be aggravated due to the inclusion of $|2\rangle\rightarrow|1\rangle$ transition frequencies, as it was observed in Ref. \cite{blok2020quantum}.

We note that it is feasible to increase the GHZ fidelity of this experiment by characterizing these cross-talks or by implementing active error mitigation techniques such as dynamical decoupling \cite{viola2005random}. The use of quantum optimal control techniques and libraries can also improve the quality of the state generated \cite{li2021pulse,ball2021software}.

Finally, the use of the 0-1 default discriminator was convenient for this experiment as the number of measurements is still feasible and the low readout error of the $|000\rangle$ state was more important to reliably reconstruct the probability amplitudes. However, we note that the protocol resources scale exponentially with the number of basis elements to be measured, so for future qutrit experiments one should program and calibrate a 0-1-2 discriminator. This task will contain several challenges, for instance, the possible overlap between IQ centroids that can lead to misclassification of several states, as shown in \cite{Qiskittutorial}, or the design of a good Kernel estimator. The implementation of a successful 0-1-2 discriminator was carried out for instance in Ref. \cite{rosenblum2018fault,blok2020quantum}.

\section{Conclusions and outlook}

We report the experimental certification of a three-qutrit maximal entangled state in a superconducting device. The generated state has a fidelity of $76 \pm 1\%$ with respect to the GHZ state, a theoretical state with these maximal entanglement properties. To achieve that, we calibrate additional qutrit gates using the pulse-level programming model Qiskit Pulse via cloud access to the IBM Quantum device \texttt{ibmq\_rome}, find an optimal circuit in terms of the number of entangling gates, and experimentally reconstruct the density matrix elements needed to obtain the fidelity with respect to the theoretical GHZ state. We report up to three independent experiments where we surpass the fidelity threshold to demonstrate the generation of a genuine three-dimensional and three-partite entangled state, proving the reproducibility of our results.

We demonstrate that superconducting quantum devices are ready to escape the flatland represented by binary quantum systems.
From a quantum technological point of view, the capacity to store more quantum information in higher dimensions has direct implications to improve quantum error-correcting codes \cite{campbell2012magic,campbell2014enhanced,raissi2018optimal,krishna2019towards}. On top of that, the multi-level quantum computation can also introduce an advantage in comparison with the current binary computation \cite{bullock2005asymptotically,gokhale2019asymptotic}. Besides applications, high-dimensional multipartite systems enable the study of the foundations of quantum physics,
where the non-locality tests such as Bell inequalities or GHZ contradictions require the use of maximally entangled states.
The non-trivial structure of the high-dimensional Hilbert space is illustrated by the fact that it took nearly 20 years to theoretically generalize the GHZ theorem beyond qubits \cite{lee2006greenberger,ryu2013greenberger,lawrence2014rotational,lawrence2017mermin}. The result required the introduction of non-Hermitian operations, also present in other tests of local realism like Bell Inequalities \cite{acin2004coincidence,alsina2016operational}. 

With this work, we show that, at the present day, anyone with internet access can experimentally study high-dimensional multipartite entanglement that heretofore was limited to a handful of laboratories worldwide. 
We anticipate an interest increase in this field with its most exciting applications yet to be imagined.

\section*{Data availability}

Experimental data can be found at \url{https://github.com/AlbaCL/GHZ-qutrit-data}.

\section*{Acknowledgements}

We acknowledge the use of IBM Quantum services for this work. We acknowledge the access to advanced services provided by the IBM Quantum Researchers Program, and also the IBM Quantum Researcher Access Award Program. M.K. and A.C.L appreciate Manuel Erhard's insights on high-dimensional entanglement.
A.A.-G. acknowledges the generous support from Google, Inc. in the form of a Google Focused Award.
This work was supported by the U.S. Department of Energy under Award No. DESC0019374 and the U.S. Office of Naval Research (ONS506661). A.A.-G. also acknowledges support from the Canada Industrial Research Chairs Program and the Canada 150 Research Chairs Program. M.K. acknowledges support from the FWF (Austrian Science Fund) via the Erwin Schr\"odinger fellowship No. J4309. 

\appendix

\section{The GHZ circuit} \label{app:GHZ_clock}

With the gates described in the main article, it was not straightforward how to find the optimal circuit to generate the GHZ state in terms of the number of gates and circuit depth that also adapts to the chip connectivity. The superposition in the $(01)$ subspace generated by the CNOT when the control qutrit is in the state $|2\rangle$ and the relative phases introduced by the $R^{(ij)}_{x}$ and $R^{(ij)}_{y}$ gates, suggested that some extra gates may be necessary to obtain ideal operations like $X^{(ij)} = |i\rangle\langle j| + |j\rangle\langle i|$. For this reason, we relied on computers to find a suitable quantum circuit.
The search for an optimal circuit could follow a standard procedure of creating a parameterized ansatz circuit that optimizes the real-valued angles of the individual gates via gradient descent or other classical optimization methods. This approach, however, may have two important disadvantages. On one side, it is not necessarily guaranteed that the obtained solution is indeed the minimal circuit. On the other side, the circuit obtained could contain an intricate parameter setting that is challenging to understand. As scientists, we aim to understand the solution so we can extract new general ideas and inspiration from it. For that reason, we followed a different approach. We restricted the angles of the rotation gates to $\theta=\pm \pi$ and $\theta=\pm \pi/2$ (except for only up to one free angle for the $R_{y}^{(01)}$ gate), which corresponding operations are easy to comprehend for humans. Then, we used the algorithm \textsc{Melvin} \cite{krenn2016automated,krenn2020computer}, adapted to operate with discrete digital qutrit gates, to find this circuit, where we also include the connectivity constraints imposed by the chip design. Indeed, we quickly discovered a feasible solution that is optimal in terms of control operations and single-qutrit gates, and which we can interpret and explain straightforwardly. 

We introduce a graphical representation to illustrate the logic of how the GHZ circuit works. The \emph{GHZ-clock} is a circle divided into three sections, each corresponding to one of the two-dimensional subspaces: $(01)$, $(12)$, and $(02)$. We draw an arrow pointing at each basis state with a length proportional to the corresponding probability amplitude. Therefore, this representation can be used to depict qutrit pure states that do not contain the three levels in one of the basis states. One may consider introducing a color or thickness dimension to the arrows to represent relative phases, although this will not be necessary for the GHZ circuit under discussion. 

As shown in Fig. \ref{fig:GHZ}, we divide each subsection into eight segments representing the computational basis states of each subspace. The circuit starts with a single arrow of unit length pointing at 12 o'clock, i.e. the $|000\rangle$ state. The first operation generates the superposition $\sqrt{2/3}|0\rangle + \sqrt{1/3}|1\rangle$ on the second qutrit, producing the state $\sqrt{2/3}|000\rangle + \sqrt{1/3}|010\rangle$, which is represented with a 2/3 arrow pointing at 12 o'clock and a 1/3 arrow pointing at the $|010\rangle$ state (approximately, 1 o'clock). The next two CNOT gates move this second arrow to the $|111\rangle$ state located in the fourth quadrant of the circle. The three single-qutrit operations that come after moving the state from the $(01)$ to the $(02)$ subspace, i.e. the $|111\rangle$ arrow moves to the third quadrant, pointing at the $|222\rangle$ state. The last part of the circuit repeats the process of generating a $|000\rangle$ and $|111\rangle$ superposition while keeping the superposition with the $|222\rangle$. This is performed by generating an equal superposition on the second qutrit, the $(|0\rangle+|1\rangle)/\sqrt{2}$ state, thus dividing into two the previous $|000\rangle$ arrow. This produces a third arrow and, therefore, equals the length of the arrows to 1/3. After applying again the two CNOT gates, this third arrow points at the $|111\rangle$ state, completing the GHZ state generation.

As stated above, this representation can be used to depict any pure state that does not contain any basis element with the three levels. It can be used for more than three particles by dividing each section into more segments. An extension that includes the remaining basis states or higher dimensions can be proposed by extending this representation to the 3D space. Similarly, one can propose a clever basis state ordering in each section that allows the derivation of simple rules that represent the application of qutrit quantum gates. These extensions are out of the scope of this work and we leave them for a future project.

\section{Genuine high-dimensional multipartite entanglement -- the high-dimensional GHZ} \label{app:entanglement}
Entanglement in multipartite and high-dimensional systems can take numerous forms. Let's consider a system of three parties. We can encode two-dimensional entanglement inside a three-dimensional space, e.g. using the 2-dimensional GHZ state $\ket{G_{i,j}}=1/\sqrt{2}(\ket{i,i,i}+\ket{j,j,j})$, where $i$ and $j$ denote the local dimension,
\begin{eqnarray}
    \rho_{2d}&=\frac{1}{3}\left(\ket{G_{0,1}}\bra{G_{0,1}}
    + \ket{G_{0,2}}\bra{G_{0,2}}
    + \ket{G_{1,2}}\bra{G_{1,2}}\right).\nonumber\\
\end{eqnarray}
Likewise, we can write 2-partite entanglement in a 3-partite system, using the 2-dimensional Bell state $\ket{B_{i,j,k}}=1/\sqrt{2}(\ket{0,0}_{i,j}+\ket{1,1}_{i,j})\ket{0}_k$, where the subscript identifies the qubit number. A 2-qubit entangled state encoded into 3 parties can be written as 
\begin{multline}
    \rho_{2n}=\frac{1}{3}\large(\ket{B_{0,1,2}}\bra{B_{0,1,2}}+\ket{B_{0,2,1}}\bra{B_{0,2,1}} \\
    + \ket{B_{1,2,0}}\bra{B_{1,2,0}}\large).
\end{multline}.

In this manuscript, we go beyond those systems and demonstrate a genuinely 3-dimensional 3-partite entangled GHZ state, $\ket{GHZ_{3d}}=1/\sqrt{3}(\ket{0,0,0}+\ket{1,1,1}+\ket{2,2,2})$. Furthermore, we demonstrate that no lower-dimensional nor lower-particle entanglement can replicate our measurement results, by employing a state-of-the-art quantum entanglement witness described in the next section.

\section{Measurement of density matrix elements and entanglement witness} \label{app:tomography}

To obtain the elements of the density matrix needed to compute the fidelity with respect to the GHZ state, we will follow the protocol introduced in Refs. \cite{bavaresco2018measurements,erhard2018experimental}. Given a three-particle density matrix of the form
\begin{equation}
    \rho = \sum_{\substack{i,j,k=0\\ l,m,n=0}}^{2}a_{ijk}a^{*}_{lmn}|ijk\rangle\langle lmn|,
    \label{eq:density_matrix}
\end{equation}
it can be shown that the real and imaginary part of an element of $\rho$ can be expressed as a combination of expectation values of $\sigma_{x}^{(ab)}$ and $\sigma_{y}^{(ab)}$ in the subspaces $(ab)=(01),(02),(12)$ as follows:
\begin{eqnarray}
\mathrm{Re}\left(\langle ijk|\rho|lmn\rangle\right) &=& \frac{1}{8} \big( \langle \sigma_{x}^{(il)}\sigma_{x}^{(jm)}\sigma_{x}^{(kn)}\rangle \nonumber \\ && - \langle \sigma_{y}^{(il)}\sigma_{y}^{(jm)}\sigma_{x}^{(kn)}\rangle \nonumber \\ &&
- \langle \sigma_{y}^{(il)}\sigma_{x}^{(jm)}\sigma_{y}^{(kn)}\rangle \nonumber \\ && - \langle \sigma_{x}^{(il)}\sigma_{y}^{(jm)}\sigma_{y}^{(kn)}\rangle \big),
\label{eq:real_part}
\end{eqnarray}
\begin{eqnarray}
\mathrm{Im}\left(\langle ijk|\rho|lmn\rangle\right) &=& \frac{1}{8}\big( \langle \sigma_{y}^{(il)}\sigma_{y}^{(jm)}\sigma_{y}^{(kn)}\rangle \nonumber \\
&&- \langle \sigma_{x}^{(il)}\sigma_{x}^{(jm)}\sigma_{y}^{(kn)}\rangle \nonumber \\
&&- \langle \sigma_{x}^{(il)}\sigma_{y}^{(jm)}\sigma_{x}^{(kn)}\rangle \nonumber \\
&&- \langle \sigma_{y}^{(il)}\sigma_{x}^{(jm)}\sigma_{x}^{(kn)}\rangle \big).
\label{eq:img_part}
\end{eqnarray}

To compute these expectation values, we will project into the $X$ and $Y$ basis and measure in the computational basis using the gates $H$ and $H_{y}$ respectively. These gates can be decomposed into the native gate set as
\begin{eqnarray}
    H^{(01)} &=& R_{x}^{(01)}\left(-\pi\right) R_{y}^{(01)}\left(\pi/2\right), \nonumber \\
    H^{(02)} &=& R_{y}^{(01)}\left(\pi\right) R_{x}^{(12)}\left(\pi\right) R_{y}^{(12)}\left(-\pi/2\right) R_{y}^{(01)}\left(-\pi\right), \nonumber \\
    H^{(12)} &=& R_{y}^{(12)}\left(-\pi/2\right)R_{x}^{(12)}\left(-\pi\right), \nonumber \\
    H_{y}^{(01)} &=& R_{x}^{(01)}\left(-\pi/2\right) R_{y}^{(01)}\left(\pi\right), \nonumber \\
    H_{y}^{(02)} &=& R_{y}^{(12)}\left(\pi\right) R_{x}^{(01)}\left(-\pi/2\right) R_{y}^{(01)}\left(\pi\right) R_{y}^{(12)}\left(-\pi\right), \nonumber \\
    H_{y}^{(12)} &=& R_{x}^{(12)}\left(-\pi/2\right) R_{y}^{(12)}\left(\pi\right).
    \label{eq:hadamard}
\end{eqnarray}

With the density matrix elements, we can compute the fidelity with respect to the GHZ state $F_\mathrm{exp}$ following Eq. \eqref{eq:fidelity}.
This equation can be derived from the general formula to compute the fidelity between two general states $F=\text{Tr}\left(\rho \sigma\right)$, where $\rho$ and $\sigma$ are the density matrices of these states. Since one of the state is a pure state, the GHZ, the density matrix $\sigma$ is simply $\sigma=|GHZ\rangle\langle GHZ|$. As a consequence, we do not need to perform a full tomography protocol.

With that fidelity, we can apply the entanglement witness protocol described in \cite{malik2016multi}. We have to show that the state generated has a multipartite entanglement structure of the form $(3,3,3)$, where $(x,y,z)$ corresponds to the Schmidt rank vector, i.e. the rank of each reduced density matrix for the three possible bipartitions. To prove that, we have to demonstrate that this state cannot be decomposed into states of lower dimensionality structure. The protocol consists of comparing the fidelity of this state and a state with $(3,3,2)$ Schmidt vector, with respect to a state with $(3,3,3)$ Schmidt vector, i.e. the GHZ state. Let's denote $F_\mathrm{max}$ as the maximal fidelity achievable by a $(3,3,2)$ state,
\begin{equation}
F_\mathrm{max} = \max_{\sigma\in(3,3,2)}\text{Tr} \left(\sigma|GHZ\rangle\langle GHZ|\right).
\end{equation}

For three-partite states, the three system bipartitions are $B_{1}\equiv A|BC$, $B_{2}\equiv AB|C$ and $B_{3}\equiv AC|B$. It can be shown, that $F_\mathrm{max}$ is bounded by
\begin{equation}
    F_\mathrm{max} \leq \min\left(\max_{\text{rank}(\sigma_{B_{k}})} \text{Tr}(\sigma |GHZ\rangle\langle GHZ|)\right),
\end{equation}
for $k=1,2,3$, where $\sigma_{B_{k}}$ is the reduced density matrix of $\sigma$ for $B_{k}$ bipartition.

Giving the Schmidt decomposition of a state across one of its bipartitions $A|\bar{A}$, $|\psi\rangle = \sum_{i=1}^{\chi}\lambda_{i}^{2}|u_{A,i}\rangle|v_{\bar{A},i}\rangle$, where $\lambda_{i}$ are the Schmidt coefficients and $\chi$ is the Schmidt rank, its maximal fidelity with another state with Schmidt rank $\xi\leq\chi$ for the same bipartition is \cite{fickler2014interface}
\begin{equation}
    F_\mathrm{max} = \max_{\text{rank}(\sigma_{\bar{A}})}\text{Tr}\left(\sigma|\psi\rangle\langle\psi|\right) = \sum_{i=1}^{\xi}\lambda_{i}^2.
\end{equation}
The GHZ Schmidt coefficients for any bipartition are $\left(1/\sqrt{3},1/\sqrt{3},1/\sqrt{3}\right)$. Thus, the maximal fidelity of state with Schmidt vector $(3,3,2)$ and a $(3,3,3)$ state like the GHZ is
\begin{equation}
    F_\mathrm{max} = \frac{1}{3} + \frac{1}{3} = \frac{2}{3}.
\end{equation}
Then, any fidelity $F_\mathrm{exp}>2/3$ can only be generated by a state with entanglement structure $(3,3,3)$, i.e. a three-dimensional and three-partite state. 

\section{Phase tracking} \label{app:track_phase}

Quantum pulse control in superconducting transmon qubits is described using the rotational frame picture. In this reference frame, the $x$ and $y$ axis rotate with the qubit frequency so Bloch sphere vectors remain stationary. When considering the dynamics of high-energy levels, one needs to include in the analysis the precision occurring in the (12) and (02) subspaces. By keeping the same rotational frame picture, the Bloch vectors in these other Bloch spheres precess with a certain angle with respect to the (01) Bloch states. This effect (that can be aggravated with other noise sources such as AC Stark effect) is effectively translated to the implementation of the qutrit gates in the (12) subspace. Instead of applying a well defined rotational gate around the $x$ or $y$ axis from that subspace, the quantum operation performed becomes a rotational gate around a unit vector $\hat{n}$ in the $(x,y)$ plane,
\begin{equation}
 R_{\hat{n}}^{(12)} = \left(
\begin{array}{ccc}
 1 & 0 & 0 \\
 0 & \cos \left(\frac{\theta }{2}\right) & e^{i\phi} \sin \left(\frac{\theta }{2}\right) \\
 0 & -e^{-i\phi} \sin \left(\frac{\theta }{2}\right) & \cos \left(\frac{\theta }{2}\right) \\
\end{array}
\right),
\label{eq:Rxy12}
\end{equation}
where gates $R_{x}^{(12)}$ and $R_{y}^{(12)}$ correspond with $\phi = \pi$ and $\phi = \pi/2$ respectively.

If we assume this behaviour in the $R^(12)$ gates that appear in the GHZ circuit, a relative phase appears in the $|222\rangle$ term. This is of no concern for the aim of this work, since the entanglement properties of a GHZ state with relative phases between the computational basis elements are not affected.

The measurement protocol that we use (projecting all basis states to the $|000\rangle$ state) requires from the $X_{+}$ gate when the basis state is $|2\rangle$. This gate can be decomposed as
\begin{equation}
    X_{+} = R_{y}^{(01)}(\pi) R_{y}^{(12)}(\pi).
    \label{eq:Xplus}
\end{equation}
Replacing $R_{y}^{(12)}$ gate by $R_{\hat{n}}^{(12)}$ from Eq. \eqref{eq:Rxy12} only introduces a global phase that is not an observable. Therefore, the measurement protocol is not affected by this implementation.

However, for the tomography protocol described in App. \ref{app:tomography} requires these $R^{(12)}$ gates and it is significantly affected by these dephasing. As an example, let's compute the probability of measuring the $|111\rangle$ state when projecting into the $x$ basis in the (12) subspace assuming that instead of $R_{y}^{(12)}$ and $R_{x}^{(12)}$ gates we are applying $R_{\hat{n}}^{(12)}$ gates. Taking the decomposition of the projective measurements from Eq. \eqref{eq:hadamard},
\begin{equation}
|\langle 111 |\sigma_{x}^{(12)}\sigma_{x}^{(12)}\sigma_{x}^{(12)}| 111\rangle|^2 = \frac{1}{24}|1 + e^{i\sum_{i=1}^{3} b_{i} - 2a_{i}}|^2,
\end{equation}
where $a_{i}$ corresponds with the phases $\phi$ introduced by the first $R_{x}^{(12)}$ gate of $H^{(12)}$, $b_{i}$ - with the phase introduced by the second rotational gate $R_{y}^{(12)}$, and the subscript indicates the qutrit. This probability oscillates between 0 and 1/6, effect observed in the oscillations shown in Fig.\ref{fig:fidelity}. 
By applying a $R_{z}^{(01)}(\varphi)$ gate between the GHZ part of the circuit and the tomography gates, the probability of this basis elements becomes
\begin{multline}
|\langle 111 |\sigma_{x}^{(12)}\sigma_{x}^{(12)}\sigma_{x}^{(12)}| 111\rangle|^2 \\
= \frac{1}{24}|1 + e^{i\left(\sum_{i=1}^{3}(b_{i} - 2a_{i}) + \varphi/2\right)}|^2.
\end{multline}
Thus, it is possible to cancel the effect of this phase accumulation by finding the proper $\varphi$.

\begin{figure*}
    \centering
    \includegraphics[width=0.9\textwidth]{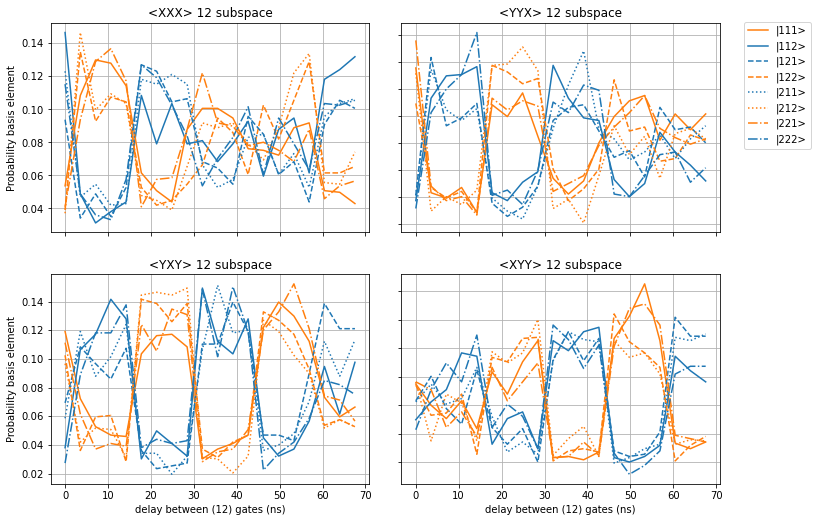}\\
    \includegraphics[width=0.9\textwidth]{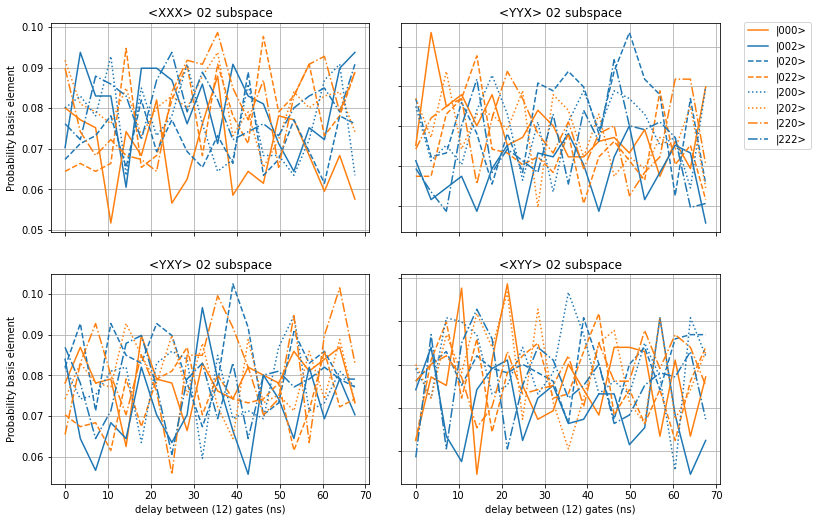}
    \caption{Oscillations of the expectation value of each basis element used to compute the tomography of a GHZ experiment as a function of the delay introduced between the two $R^{(12)}$ gates used in the projective measurements. This experiment was carried out using the \texttt{ibmq\_rome} chip on transmons Q2-Q3-Q4 on a different day than the results shown in the main paper. The fidelities obtained are $F_\mathrm{exp}=0.73 \pm 0.01$ and $F_\mathrm{raw}=0.67\pm 0.01$.}
    \label{fig:delays}
\end{figure*}

Another strategy to cancel this dephasing consist of introducing a delay between the implementation of the $R^{(12)}$ gates in the tomography part of the circuit. Figure \ref{fig:delays} shows the result of introducing these delays. The experiment was carried out on a different day than the results presented in the main paper, and also on the \texttt{ibmq\_rome} device transmons Q4, Q3 and Q2, and delivered the fidelity of $F_\mathrm{exp}=0.73 \pm 0.01$ and $F_\mathrm{raw}=0.67\pm 0.01$. The effect of introducing these delays is evident in the (12) subspace but indistinguishable from noise in the (02) subspace.

\section{Data and error analysis} \label{app:error_data}

To reconstruct each density matrix element, we have to compute the expectation value of eight Pauli strings, see Eqs.~(\ref{eq:real_part}--\ref{eq:img_part}). To measure each expectation value, we project into the corresponding Pauli basis and measure on the computational basis. As described in the main text, our measurement protocol discriminates between the state $|000\rangle$ and the rest of the possible states. Thus, we will measure the amplitudes of the computational basis elements by local-transforming each of them to the $|000\rangle$ state. For instance, to obtain the $|011\rangle$ amplitude we will apply a $X^{(01)}$ gate on the second and third qutrit, while for obtaining the amplitude of the $|020\rangle$ state we will apply the $X^{(01)}X^{(12)}$ transformation on the second qutrit, i.e. the operation described in Eq.\eqref{eq:Xplus}. Since we are only interested in density matrix elements that involve up to two levels, each expectation value will only require a two-dimensional measurement, either between the $(01)$, $(12)$ or $(02)$ levels, so we will have to compute the eight possible computational basis states for each level pair. This measurement protocol implies that we need to perform eight experiments to obtain the expectation value of each Pauli string. 

We assume that each experiment is independent and it takes approximately the same amount of time since the only difference between them is the application of up to two single-qutrit gates. We assume the counts obtained follow a Binomial distribution with $\mu = np$, $\text{Var}=np(1-p)$ with $p=n_\mathrm{counts}/n$ and $n$ being the number of shots per experiment. Since $n\gg 1$, this distribution can be approximated as a Normal distribution.

To obtain each expectation value, we assign the eigenvalue $+1$ or $-1$ (eigenvalues of $\sigma_{x,y}^{(ij)}$) to each output. Then, we add them following Eq. \eqref{eq:real_part} and Eq. \eqref{eq:img_part} to reconstruct the density matrix element. We will compute and propagate the errors according to
\begin{equation}
    \sigma^{(ij)}_{abc} = \sqrt{\sum_{k=1}^{8}\text{Var}(\langle abc\rangle_{k}^{(ij)})}, 
\end{equation}
where $k$ represents the eight computational basis elements, $abc$ corresponds to the Pauli string (e.g. $xxy\equiv$ $\sigma_{x}\sigma_{x}\sigma_{y}$) and $(ij)$ the two levels. Thus, the standard deviation of the real part of a density matrix element $\langle iii|\rho|jjj\rangle$ will be
\begin{equation}
    \sigma_{\text{Re}}^{(ij)} = \frac{1}{8} \sqrt{\left(\sigma^{(ij)}_{xxx}\right)^{2}+ \left(\sigma^{(ij)}_{yyx}\right)^{2}+ \left(\sigma^{(ij)}_{yxy}\right)^{2} + \left(\sigma^{(ij)}_{xyy}\right)^{2}},
    \label{eq:SD_real}
\end{equation}
and similarly for the imaginary part.
For the diagonal matrix elements, the standard deviation is just the square root of the variance for each number of diagonal element counts.

To obtain the fidelity with respect to the original GHZ state, we use the expression from Eq. \eqref{eq:fidelity} and propagate the errors accordingly,
\begin{widetext}
\begin{equation}
    \sigma_{F} = \frac{1}{3}\sqrt{\sum_{i=0}^{2}\left(\sigma^{(ii)}\right)^2 + \left(\frac{1}{8}\right)^2\left(4\left(\sigma_{\text{Re}}^{(01)}\right)^2 + 4\left(\sigma_{\text{Re}}^{(12)}\right)^2 + 4\left(\sigma_{\text{Re}}^{(02)}\right)^2\right)}.
\end{equation}
\end{widetext}
Notice that due to the Hermiticity of the density matrix and because the original GHZ does not contain complex coefficients, we can compute the fidelity only with the real part of the density matrix elements by using the identity $z + z^{*} = 2 \ \text{Re}(z)$.

Table \ref{tab:data} shows the mitigated expectation values necessary to reconstruct the real part of the density matrix and compute the fidelity of the state with respect to the GHZ state presented in the main paper. The real part of the density matrix $\tilde{\rho}$ needed to obtain the fidelity is 
\begin{equation}
    \text{Re}(\tilde{\rho}) = \left(
    \begin{array}{ccc}
0.29\pm 0.01 & 0.282\pm 0.006 & 0.165\pm 0.006 \\
0.282\pm 0.006 & 0.36 \pm 0.02  & 0.214 \pm 0.006 \\
0.165\pm 0.006 & 0.214 \pm 0.006 & 0.31 \pm 0.02 
    \end{array}
    \right),
    \label{eq:density_matrix_real}
\end{equation}
where we use the tilde to distinguish it from the total density matrix of the generated state.

\begin{table}[t!]
    \centering
    \begin{tabular}{c|c|c|c}
         Pauli string &  (01) subspace & (12) subspace & (02) subspace \\
         \hline
         \hline
         $\langle \sigma_{x}\sigma_{x}\sigma_{x}\rangle$
         & $0.59 \pm 0.02$ & $0.43 \pm 0.02$ & $0.34 \pm 0.02$  \\
         \hline
         $\langle \sigma_{y}\sigma_{y}\sigma_{x}\rangle$
         & $-0.58 \pm 0.02$ & $-0.43 \pm 0.02$ & $-0.32 \pm 0.02$  \\
         \hline
         $\langle \sigma_{y}\sigma_{x}\sigma_{y}\rangle$
         & $-0.57 \pm 0.02$ & $-0.43 \pm 0.02$ & $ -0.34 \pm 0.02$  \\
         \hline
         $\langle \sigma_{x}\sigma_{y}\sigma_{y}\rangle$
         & $ -0.53 \pm 0.02$ & $-0.43 \pm 0.02$ & $-0.32 \pm 0.02 $  \\
         \hline
    \end{tabular}
    \caption{Expectation values of the Pauli strings necessary to reconstruct the real parts of the density matrix elements following the tomography protocol presented in Eq. \eqref{eq:real_part}. The total number of shots per experiment is $n=1024$. To compute each expectation value, one needs to perform a total of 8 experiments in order to measure the probability amplitudes of each computational basis state following the measurement protocol presented in the main text. We apply the measurement error mitigation matrix from Fig.~\ref{fig:mit} as described in the main text.}
    \label{tab:data}
\end{table}

\section{Measuring the relative phases} \label{app:relative_phases}

To compute the fidelity with respect to the original GHZ state, one only needs the real part of the density matrix elements. However, we can also measure the imaginary parts of the density matrix and obtain the relative phases between the basis elements (if any).

\begin{table}[t!]
    \centering
    \begin{tabular}{c|c|c|c}
         Pauli string &  (01) subspace & (12) subspace & (02) subspace \\
         \hline
         \hline
         $\langle \sigma_{x}\sigma_{x}\sigma_{x}\rangle$
         & $0.69 \pm 0.03$ & $0.41 \pm 0.03$ & $0.46 \pm 0.03$  \\
         \hline
         $\langle \sigma_{y}\sigma_{y}\sigma_{x}\rangle$
         & $-0.68 \pm 0.03$ & $-0.35 \pm 0.03$ & $-0.47 \pm 0.03$  \\
         \hline
         $\langle \sigma_{y}\sigma_{x}\sigma_{y}\rangle$
         & $-0.58 \pm 0.03$ & $-0.36 \pm 0.03$ & $-0.46 \pm 0.03$  \\
         \hline
         $\langle \sigma_{x}\sigma_{y}\sigma_{y}\rangle$
         & $-0.67 \pm 0.03$ & $-0.37 \pm 0.03$ & $-0.46 \pm 0.03$  \\
         \hline
         \hline
        $\langle \sigma_{y}\sigma_{y}\sigma_{y}\rangle$
         & $0.03 \pm 0.03$ & $0.12 \pm 0.03$ & $-0.06 \pm 0.03$  \\
         \hline
         $\langle \sigma_{x}\sigma_{x}\sigma_{y}\rangle$
         & $-0.006 \pm 0.034$ & $-0.13 \pm 0.03$ & $0.1 \pm 0.03$  \\
         \hline
         $\langle \sigma_{x}\sigma_{y}\sigma_{x}\rangle$
         & $-0.006 \pm 0.033$ & $-0.15 \pm 0.03$ & $0.16 \pm 0.03$  \\
         \hline
         $\langle \sigma_{y}\sigma_{x}\sigma_{x}\rangle$
         & $-0.07 \pm 0.03$ & $-0.15 \pm 0.03$ & $0.16 \pm 0.03$  \\
         \hline
    \end{tabular}
    \caption{Expectation values of the Pauli strings necessary to reconstruct the real and imaginary parts of the density matrix elements following the tomography protocol presented in Eq. \eqref{eq:real_part} and \eqref{eq:img_part}. The total number of shots per experiment is $n=512$. This experiment was performed to compute the relative phases of the GHZ state and is independent from the results shown in the main paper.
    }
    \label{tab:data_phases}
\end{table}

\begin{table*}
\centering
 \begin{tabular}{l  c  c  c  c c  c  c  c} 
  & & & & $Q_0$ & $Q_1$ & $Q_2$ & $Q_3$ & $Q_4$ \\ 
 \hline
 \hline
 \multicolumn{4}{l}{Qutrit $|0\rangle\leftrightarrow|1\rangle$ frequency, $\omega_{01}/2\pi$ (GHz)}& 4.969 & 4.770 & 5.015 & 5.259 & 4.998\\ 
 \hline
  \multicolumn{4}{l}{Qutrit $|1\rangle\leftrightarrow|2\rangle$ frequency, $\omega_{12}/2\pi$ (GHz)} & 4.631 & 4.443 & 4.677 & 4.926 & 4.658 \\
 \hline
 \multicolumn{4}{l}{Lifetime $T_1^{|1\rangle\rightarrow|0\rangle}$ ($\mu$s)} & 110 & 98 & 36 & 118 & 74 \\ 
 \hline
 \multicolumn{4}{l}{Echo time $T_{2\textrm{Echo}}$, $|1\rangle$/$|0\rangle$ ($\mu$s)} & 76 & 59 & 69 & 133 & 94 \\ 
 \hline
 \multicolumn{4}{l}{Readout error $\left(10^{-2}\right)$} & 2.5 & 4.3 & 2.6 & 2.1 & 2.0 \\ 
 \hline
 \multicolumn{4}{l}{Prob. Prep. \(|0\rangle\) Meas. \(|1\rangle\) $\left(10^{-2}\right)$} & 3.0 & 5.2 & 3.5 & 3.5 & 2.9 \\ 
 \hline
 \multicolumn{4}{l}{Prob. Prep. \(|1\rangle\) Meas. \(|0\rangle\) $\left(10^{-2}\right)$} & 1.9 & 3.4 & 1.7 & 0.8 & 1.0 \\ 
 \hline
 \multicolumn{4}{l}{$X$ gate error $\left(10^{-4}\right)$} & 2.5 & 3.0 & 3.5 & 3.0 & 4.2 \\
 \hline 
 \multicolumn{4}{l}{$u2$ gate duration (ns)} & 36 & 36 & 36 & 36 & 36\\
 \hline 
 &&&&&&&& \\ 
 & $[0,1]$ & $[1,0]$ & $[1,2]$ & $[2,1]$ & $[2,3]$ & $[3,2]$ & $[3,4]$ & $[4,3]$ \\ 
 \hline
 \hline
 CNOT gate error $\left(10^{-3}\right)$ & 6.5 & 6.5 & 18.6 & 18.6 & 8.8 & 8.8 & 10.7 & 10.7\\
 \hline 
 CNOT gate duration (ns) & 320 & 356 & 1109 & 1145 & 377 & 341 & 476 & 512\\
 \hline 
\end{tabular}
 \caption{Calibration data for \texttt{ibmq\_rome} for the experimental results shown in the main paper. Note: $u2$ stands for one of the default IBM's single-qubit gate while $U2$ from Fig.2 from the main text, stands for the transmon channel number two.
 }
 \label{Table:chip_properties} 
\end{table*}

\begin{figure*}
    \centering
    \includegraphics[width=0.33\linewidth]{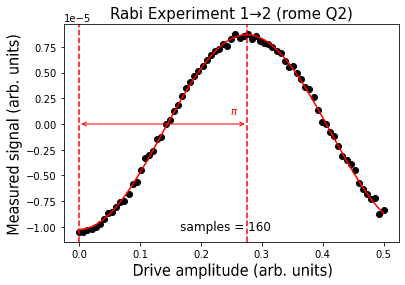}%
    \includegraphics[width=0.33\linewidth]{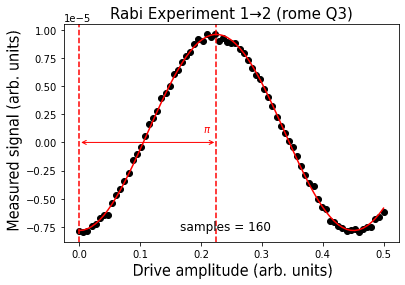}%
    \includegraphics[width=0.315\linewidth]{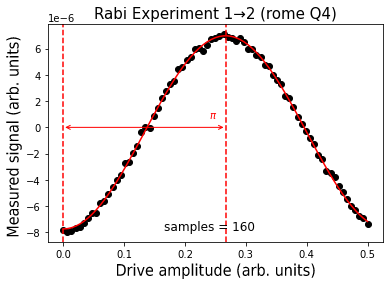}
    \caption{Rabi $1\rightarrow 2$ experiments performed to calibrate the amplitude of the $\pi_{1\rightarrow 2}$ pulse. The amplitudes obtained are 0.278, 0.225, 0.267 for transmons Q2, Q3 and Q4 respectively.}
    \label{fig:rabi_drag}
\end{figure*}

\begin{figure}[t!]
   \centering
\includegraphics[width=\columnwidth]{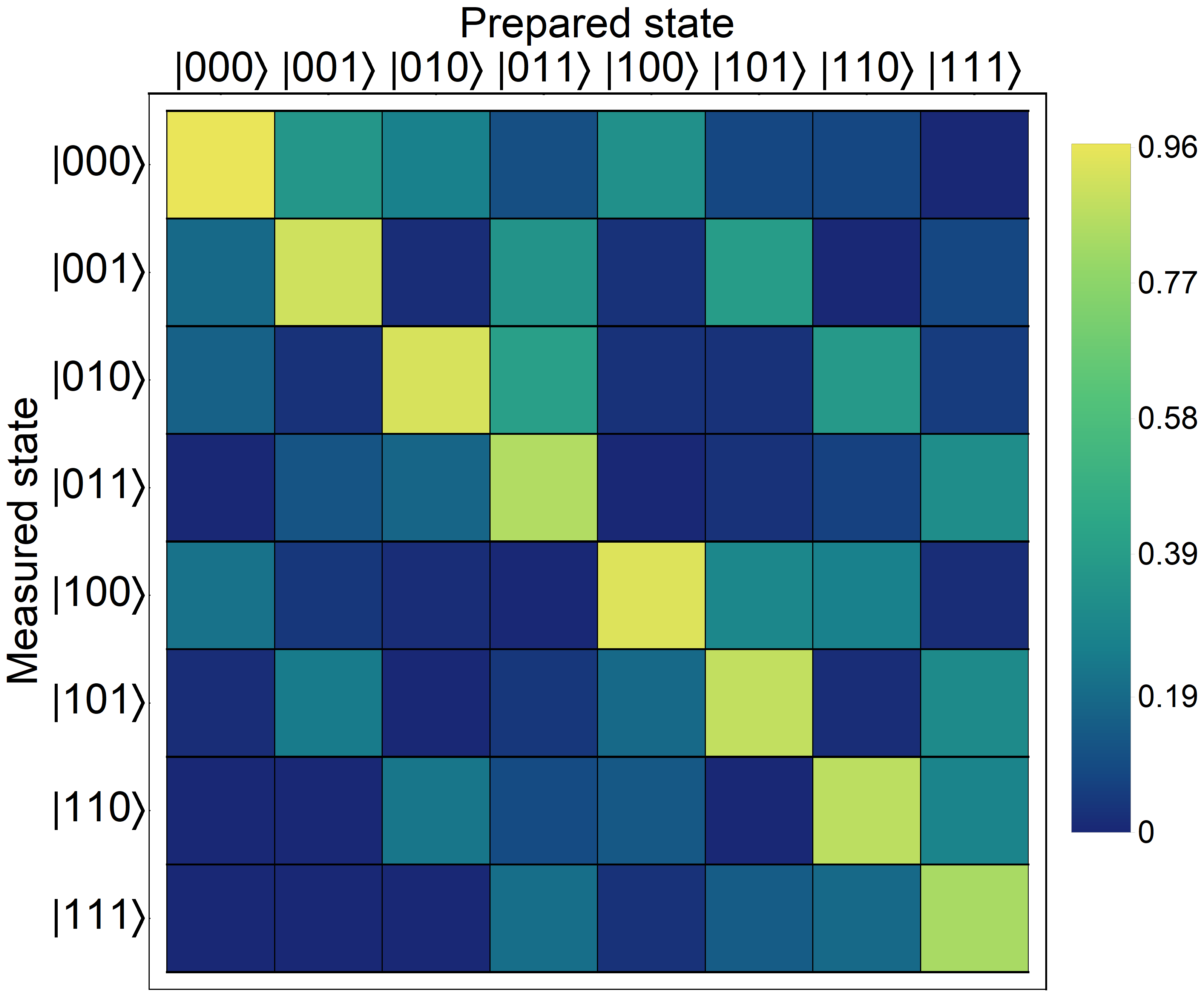}
   \caption{Calibration matrix for \(|0\rangle\)/\(|1\rangle\) readout on the 
   [Q4, Q3, Q2] subset of qutrits of the \texttt{ibmq\_rome} chip.}
    \label{fig:mit}
\end{figure}

Although the original GHZ state \cite{greenberger1989going} does not contain relative phases between the computational basis states, any state of the form
\begin{equation}
    |GHZ\rangle = \frac{1}{\sqrt{3}}\left(|000\rangle + e^{i\phi_{1}} |111\rangle + e^{i\phi_{2}}|222\rangle\right),
    \label{eq:GHZphases}
\end{equation}
with $-\pi \leq \phi_{1},\phi_{2}\leq \pi$ share the same entanglement properties as the original GHZ state ($\phi_{1}=\phi_{2}=0$), i.e. Schmidt vector of $(3,3,3)$ \cite{huber2013structure} and maximal entropy for all bipartitions. We can obtain the values of these relative phases by comparing the experimental density matrix with the theoretical one. The partial GHZ density matrix with phases is
\begin{equation}
    \tilde{\rho}_{GHZ}\left(\phi_{1},\phi_{2}\right) = \frac{1}{3}\left(
    \begin{array}{ccc}
         1 & e^{-i\phi_{1}} & e^{-i\phi_{2}}  \\
         e^{i\phi_{1}} & 1 & e^{i\left(\phi_{1}-\phi_{2}\right)} \\
         e^{-i\left(\phi_{1}-\phi_{2}\right)} & e^{i\phi_{2}} & 1
    \end{array}
    \right),
    \label{eq:density_matrix_theory}
\end{equation}
where we used again $\tilde{\rho}$ to distinguish it from the total density matrix $\rho$. From this matrix we compute $\text{Arg}\langle 000|\rho|111\rangle=-\phi_{1}$ and $\text{Arg}\langle 000|\rho|222\rangle=-\phi_{2}$. Similarly, we can obtain the difference between the phases using the third off-diagonal term $\text{Arg}\langle 111|\rho|222\rangle=\phi_{1}-\phi_{2}$ and compare it with the difference between the individually obtained phases.

We performed another experiment that measures both the real and imaginary parts of the GHZ state. The experiment was performed on the \texttt{ibmq\_rome} device using transmons Q1, Q2 and Q3 with reported fidelities of $F_\mathrm{exp} = 0.78 \pm 0.01$ and $F_\mathrm{raw} = 0.69 \pm 0.02$.
Table \ref{tab:data_phases} show the expectation values needed to reconstruct the density matrix elements $\langle 000|\rho|111\rangle$, $\langle 000|\rho|222\rangle$ and $\langle 111|\rho|222\rangle$. The total number of shots for each experiment is $n=512$.

The corresponding density matrix elements obtained are
\begin{widetext}
\begin{equation}
\tilde{\rho} = \left(\begin{array}{ccc}
 0.35\pm 0.02 & (0.327\pm0.008) + i(0.014 \pm 0.008) & (0.231\pm 0.008) -i(0.059\pm 0.008) \\
 (0.327\pm 0.008) - i(0.014\pm 0.008) & 0.29\pm 0.02 & (0.186\pm 0.007) + i(0.070\pm 0.008) \\
 (0.231\pm 0.008) + i(0.059\pm 0.008) & (0.186\pm 0.007) - i(0.070\pm 0.008) & 0.21\pm 0.02 \\
\end{array}
\right),
\label{eq:partial_matrix}
\end{equation}
\end{widetext}
where we used $\tilde{\rho}$ to distinguish it from the total density matrix $\rho$.

The experimental $\tilde{\rho}$ matrix written in terms of the absolute and argument values of each measured density matrix element is
\begin{widetext}
\begin{equation}
    \tilde{\rho} = \left(
    \begin{array}{ccc}
0.35\pm 0.02 & (0.327\pm 0.008)e^{i(0.04\pm 0.03)} & (0.239\pm 0.008)e^{-i(0.25\pm 0.03)} \\
(0.327\pm 0.008)e^{-i(0.04\pm 0.03)} & 0.29\pm0.02  & (0.199\pm 0.007 )e^{i(0.36\pm 0.04)} \\
(0.239\pm 0.008)e^{i(0.25\pm 0.03)} & (0.199\pm 0.007)e^{-i(0.36\pm 0.04)} & 0.21\pm 0.02 
    \end{array}
    \right).
    \label{eq:density_matrix_abs_arg}
\end{equation}
\end{widetext}
To obtain the corresponding errors we apply the standard error propagation formula
\begin{equation}
    \sigma_{f(x,y)}=\sqrt{\left(\frac{\partial f}{\partial x}\right)^2\sigma_{x}^2 + \left(\frac{\partial f}{\partial y}\right)^2\sigma_{y}^2}, 
\end{equation}
where $x$ and $y$ are the real and imaginary parts of each element, $\sigma_{x}$ and $\sigma_{y}$ their standard deviations as computed in Eq. \ref{eq:SD_real}, and $f(x,y) = \sqrt{x^2 + y^2}$, for the absolute value, and $f(x,y)=2\arctan(y/(\sqrt{x^2 + y^2} + x)$ for the argument.

From \eqref{eq:density_matrix_abs_arg}, we obtain the relative phases
\begin{eqnarray}
 \phi_{1} &=& -0.04 \pm 0.03, \nonumber \\
 \phi_{2} &=& 0.25 \pm 0.03,
\end{eqnarray}
and $(\phi_1-\phi_2) = -0.36 \pm 0.04$, estimated from the $\langle 111|\rho|222\rangle$ term, and  $(\phi_1-\phi_2) = -0.29 \pm 0.04$, estimated from the above values of the phases.

\section{Chip characterization and readout calibration matrix}
\label{app:chip}

In this paper we used \texttt{ibmq\_rome}, which is one of the IBM Quantum Falcon Processors. Pulse-level control was implemented using Qiskit Pulse~\cite{alexander2020qiskit} to generate the pulses for the single qutrit gates $R_{x}^{(12)}$ and $R_{y}^{(12)}$ at the $f_{12}$ transition frequency. The device specifications were provided by IBM Quantum~\cite{ibmq_systems} and are shown in Table~\ref{Table:chip_properties}. We used qutrits Q4-Q3-Q2 for the results shown in the main paper, also denoted as $[4,3,2]$ subset.

For each of these transmons, we performed a Rabi experiment to obtain the amplitude of the $\pi_{1\rightarrow 2}$ pulse used to define the rotational gates in the (12) subspace. We also calibrated the DRAG pulses. Figure \ref{fig:rabi_drag} shows the plots of this calibration.

Measurement error mitigation was performed by correcting the average shot counts collected in the experiments by using the calibration matrix from Fig. \ref{fig:mit} obtained using the Qiskit Ignis framework. The calibration matrix was generated by preparing 8 basis input states and computing the probabilities of measuring counts in all other basis states.

\bibliography{ref}

\end{document}